\documentclass{PoS}
\usepackage{epsfig}
\usepackage{bm}
\usepackage{amssymb}
\usepackage{fontenc}
\usepackage{times}
\usepackage{mathptmx}
\usepackage{graphicx}

\title{QCD thermodynamics with colour-sextet quarks}

\ShortTitle{QCD thermodynamics with colour-sextet quarks}

\author{\speaker{D.~K.~Sinclair}%
\thanks{This research was supported in part by US Department of Energy
contract DE-AC02-06CH11357, and in part under a Joint Theory Institute
(JTI) grant.}\\
        HEP Division and JTI, Argonne National Laboratory, 9700 South Cass Ave.,
        Argonne, IL, 60439, USA 
        \\
        E-mail: \email{dks@hep.anl.gov}}

\author{J.~B.~Kogut
\thanks{Supported in part by a National Science Foundation grant 
NSF PHY03-04252.}\\
Department of Energy, Division of High Energy Physics, Washington, DC 20585,
USA\\
and\\
Dept. of Physics -- TQHN, Univ. of Maryland, 82 Regents Dr., College
Park, MD 20742, USA\\
        E-mail: \email{jbkogut@umd.edu}}

\abstract{We study QCD with two flavours of colour-sextet quarks as a candidate
walking-Technicolor theory. We simulate lattice QCD with two flavours of 
colour-sextet staggered quarks at finite temperatures to observe the scales
of confinement and chiral-symmetry breaking. These should give us some 
indication as to whether the massless theory has an infrared fixed point making
it a conformal field theory, or whether it exhibits confinement and chiral
symmetry breaking with a slowly varying coupling constant, i.e. `walks'. We
find that unlike the case with fundamental quarks, the deconfinement and
chiral-symmetry restoration transitions are far apart. The values of 
$\beta=6/g^2$ for both transitions increase when $Ta$ is decreased from
$\frac{1}{4}$ to $\frac{1}{6}$ as would be expected for finite temperature
transitions of an asymptotically-free field theory. So far we see no suggestion
of conformal behaviour.}

\FullConference{The XXVII International Symposium on Lattice Field Theory\\
		 July 26-31, 2009\\
		 Peking University, Beijing, China}

\begin{document}

\section{Introduction}

We are interested in extensions of the standard model which have a 
strongly-interacting Higgs sector. The most popular of such theories are known
as Technicolor theories \cite{Weinberg:1979bn,Susskind:1978ms}. 
These are QCD-like theories in which the techni-pions
give mass to the $W$ and $Z$ through the Higgs mechanism. Such theories need to
be extended in order to also give masses to the quarks and leptons. 
Phenomenological difficulties with such theories can be avoided if the running
coupling constant `walks', that is evolves very slowly. Such theories are called
Walking Technicolor models 
\cite{Holdom:1981rm,Yamawaki:1985zg,Akiba:1985rr,Appelquist:1986an}.

The evolution of the coupling constant $g$ in QCD-like theories is described by
$\beta(g)$ defined by
\begin{equation}
\beta(g) = \mu{\partial g \over \partial \mu} =
- \beta_0{g^3 \over (4\pi)^2} - \beta_1{g^5 \over (4\pi)^4}...
\end{equation}
where $\mu$ is the momentum scale at which the running coupling constant
$g(\mu)$ is defined. $\beta_0$,$\beta_1$... are given by perturbation theory.
If the number of flavours $N_f$ is small enough, $\beta_0, \beta_1 > 0$ and
the theory is asymptotically free and confining. Chiral symmetry is 
spontaneously broken. For $N_f$ sufficiently large $\beta_0 < 0$ and asymptotic
freedom and confinement are lost. Between these extremes, there is a range of 
$N_f$ for which $\beta_0 > 0$, $\beta_1 < 0$. If these 2 terms describe the 
physics, there is a second fixed point which is infrared attractive. The 
massless theory is then conformally invariant. There is, however, an alternative
scenario; the coupling constant could become large enough that a chiral
condensate forms before the would-be infrared (IR) fixed point is reached. This
reduces the screening of the colour `charge', and the coupling increases again, 
giving  confinement. Near the would-be IR fixed point the coupling walks.
Perturbation theory cannot determine whether the theory is conformal or walking.
Lattice gauge theory simulations provide the only reliable method for studying
these non-perturbative questions.

The advent of the LHC and increased computing power have revived interest in
using lattice gauge theory simulations to study candidate conformal and
walking gauge theories. A number of studies have been made of QCD with $N_f$
flavours of fundamental quarks, with $N_f$ large (as large as $17)$ 
\cite{Kogut:1985pp,Fukugita:1987mb,Ohta:1991zi,Kim:1992pk,Brown:1992fz,
Iwasaki:1991mr,Iwasaki:2003de,Deuzeman:2008sc,Deuzeman:2009mh,
Appelquist:2009ty,Appelquist:2007hu,Jin:2008rc,Jin:2009,Fodor:2009wk}. There 
have also been studies of QCD with $N_f=2$ color-sextet (symmetric tensor) 
quarks \cite{Shamir:2008pb,DeGrand:2008kx,Fodor:2008hm}, and of 2-colour QCD 
with $N_f=2$ adjoint fermions 
\cite{Catterall:2007yx,Catterall:2008qk,DelDebbio:2008zf,DelDebbio:2009fd,%
Hietanen:2009az}.

We are studying a particular candidate theory, QCD with 2 colour-sextet quarks.
On the lattice, we are using unimproved staggered quarks and a simple Wilson
plaquette gauge action for our simulations. Exact RHMC simulations are used to 
tune to 2 quark flavours.

For QCD with sextet quarks, asymptotic freedom is lost at $N_f=3\frac{3}{10}$.
$\beta_1$ changes sign at $N_f=1\frac{28}{125}$. A rainbow graph approximation
predicts that a condensate forms for $N_f<2\frac{163}{325}$. (See
\cite{Dietrich:2006cm} for a summary of such perturbative results for $SU(N)$
For recent estimates of this boundary using other techniques, see for example
\cite{Poppitz:2009uq,Armoni:2009jn}).
However, preliminary lattice results of DeGrand, Shamir and Svetitsky using
Wilson quarks suggest $N_f=2$ is conformal \cite{Shamir:2008pb}.

We are simulating thermodynamics of this theory to better understand how
confinement and chiral symmetry are realized. Low statistics lattice studies of
the same theory, using improved Wilson quarks have been performed by DeGrand, 
Shamir and Svetitsky \cite{DeGrand:2008kx}. 
Whereas they find coincident deconfinement and 
chiral-symmetry restoration transitions, we find that the chiral-symmetry
restoration transition occurs at a much higher temperature than the 
deconfinement transition. Both transitions appear to be finite temperature
transitions and not bulk transitions. After this we intend to simulate at zero 
temperature to further clarify whether it is a conformal field theory, or if it 
walks.

\section{Simulations and results}

We are simulating the thermodynamics of QCD with 2 sextet-quark flavours
on $8^3 \times 4$, $12^3 \times 4$ and
$12^3 \times 6$ lattices. On each lattice we run at quark masses $m=0.02$,
$m=0.01$ and $m=0.005$ in an attempt to access the chiral limit. We run at
a number of couplings in the range $5.0 \le \beta=6/g^2 \le 7.0$ to cover both
the expected transitions (as well as one unexpected transition). For each
$(\beta,m)$ we run for a minimum of 10,000 length-1 trajectories, increasing
this to 50,000-100,000 trajectories close to the deconfinement and additional
non-chiral transitions. In addition we used different starts in the deconfined
region to study the remnants of the $Z_3$ centre symmetry. The deconfinement
transition is identified by an abrupt increase in the colour-triplet Wilson
Line (Polyakov Loop). The chiral phase transition occurs where the chiral
condensate ($\langle\bar{\psi}\psi\rangle$) vanishes in the massless limit.

For our $N_t=4$ runs, we find consistency between our $8^3 \times 4$ and
$12^3 \times 4$ simulations for the masses we have considered, indicating that
finite size effects are small. We therefore present only our $12^3 \times 4$ 
measurements.

\begin{figure}[htb]
\parbox{2.9in}{
\epsfxsize=2.9in
\epsffile{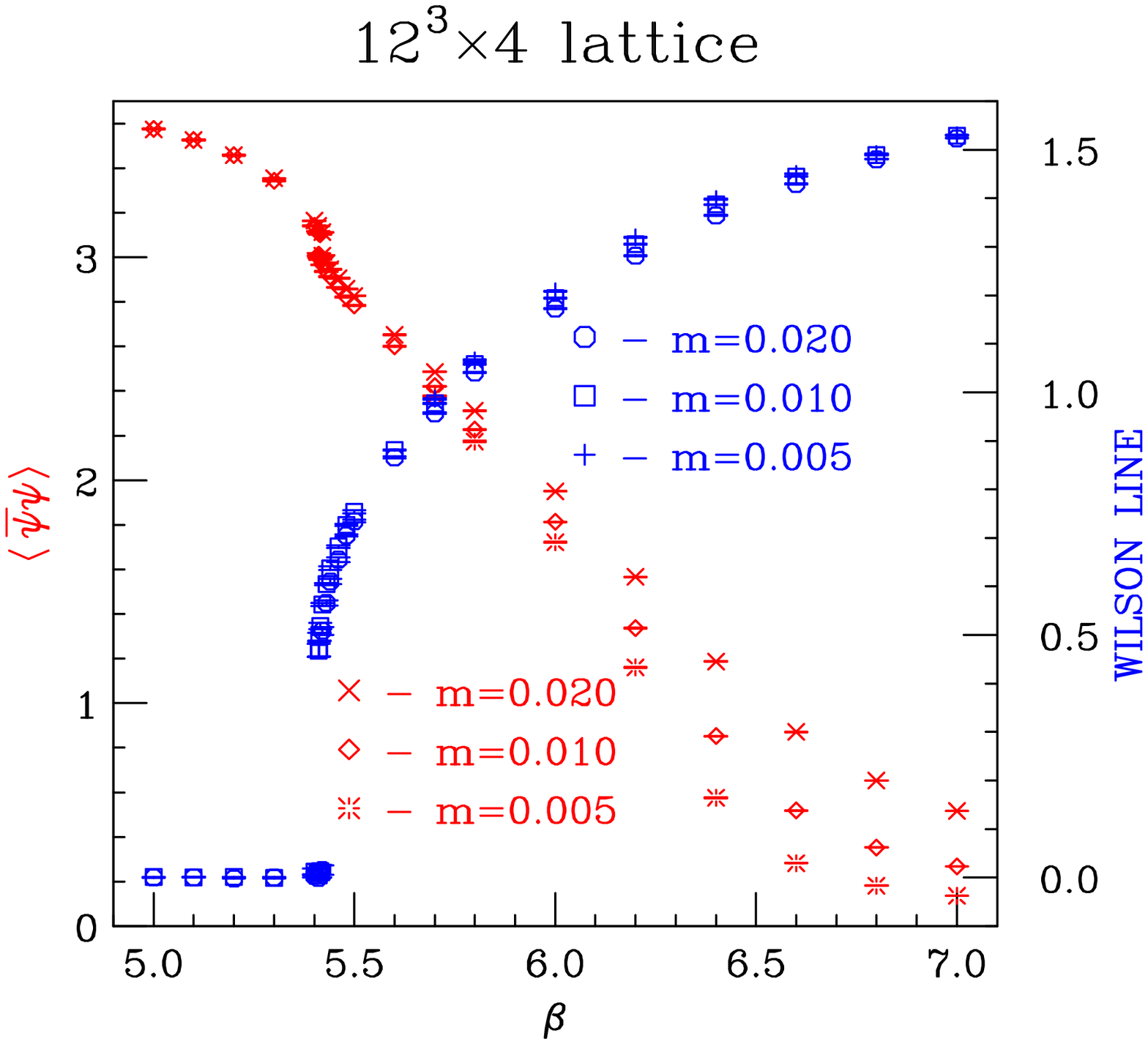}
\caption{Wilson Lines and chiral condensates on a $12^3 \times 4$ lattice as 
functions of $\beta=6/g^2$.}
\label{fig:wil-psi_12x4}
}
\parbox{0.2in}{}
\parbox{2.9in}{
\epsfxsize=2.9in 
\epsffile{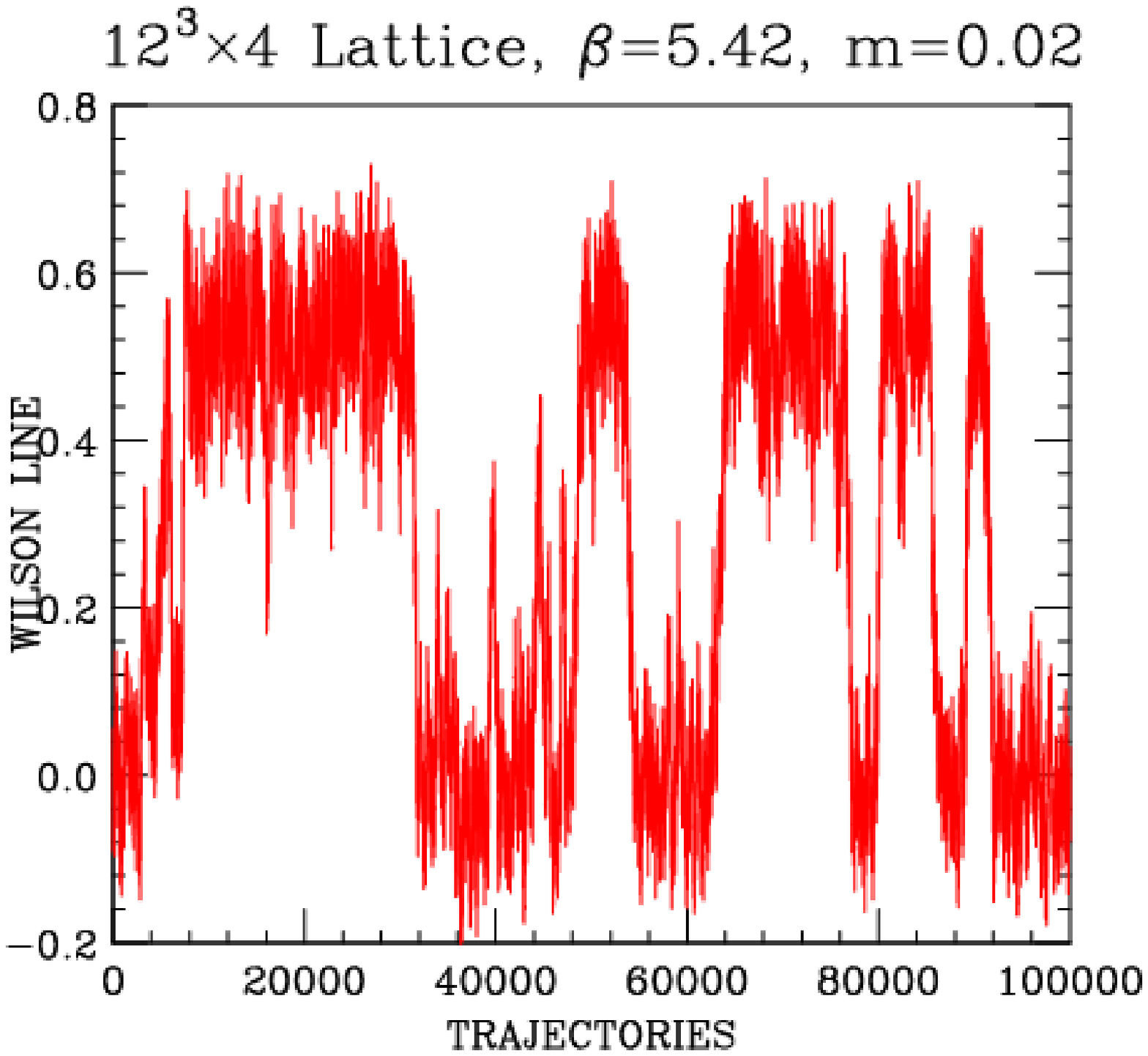}
\caption{Evolution of the triplet Wilson line as a function of trajectory
number on a $12^3 \times 4$ lattice at $\beta=5.42$, $m=0.02$.}
}
\label{fig:wil-5.42}
\end{figure}

Figure~\ref{fig:wil-psi_12x4} shows the colour-triplet Wilson Line(Polyakov
Loop) and the chiral condensate($\langle\bar{\psi}\psi\rangle$) as functions of
$\beta=6/g^2$, for each of the 3 quark masses on a $12^3 \times 4$ lattice.
There is clearly an abrupt transition in the Wilson Line for $\beta$ just 
above $5.4$. Figure~\ref{fig:wil-5.42} shows the `time' evolution of the 
triplet Wilson Line at $\beta=5.42$, $m=0.02$. The apparent 2-state signal, 
which is born out by histogramming, allows us to conclude that deconfinement
occurs at $\beta=5.420(5)$ for $m=0.02$ and $N_t=4$. Similarly we find that
the deconfinement transition for $m=0.01$ is at $\beta=5.412(1)$. Because the
chiral-symmetry restoration phase transition is only expected for massless
quarks, it is less easy to pinpoint. Our estimate for the position of this
transition is at $\beta \approx 6.5$.

Above the deconfinement transition, in addition to the state where the Wilson
Line is positive, there are additional states in which the Wilson line is 
oriented in the direction of one of the non-trivial cube roots of unity, the
remnant of the $Z_3$ centre symmetry. However, these states are only metastable,
eventually decaying into the state with a real positive Wilson Line. The
lifetimes of these states increase as $\beta$ is increased away from the
transition, so that on a $12^3 \times 4$ lattice, we have yet to observe such a
decay beyond $\beta=5.46$. Above $\beta \approx 5.9$ these states with complex
Wilson Lines transition to states with negative Wilson Lines.

\begin{figure}[htb]
\epsfxsize=3in
\centerline{\epsffile{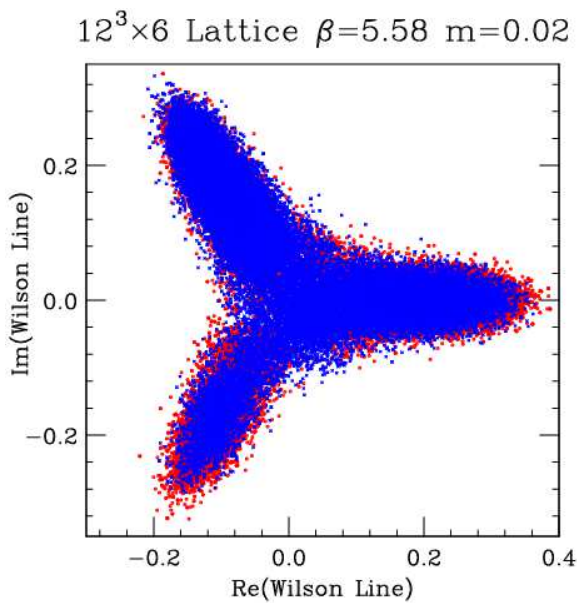}}
\caption{Scatterplot of triplet Wilson Lines at $\beta=5.58$, $m=0.02$, in the 
deconfined regime on a $12^3 \times 6$ lattice.} 
\label{fig:wil-5.58}
\end{figure}

Now let us turn our attention to $N_t=6$. On the $12^3 \times 6$ lattices, the
$Z_3$ centre symmetry is again manifest, above the deconfinement transition,
and the Wilson Line shows a clear 3-state signal. Figure~\ref{fig:wil-5.58}
demonstrates this at $\beta=5.58$, $m=0.02$, which is above this transition. 
This graph is a scatterplot of the Wilson Lines for 100,000 trajectories.
Tunnelings occur between these 3 states in all 6 directions, implying that all
3 states are stable, in contrast to what we saw for $N_t=4$.

\begin{figure}[htb]
\parbox{2.9in}{
\epsfxsize=2.9in
\epsffile{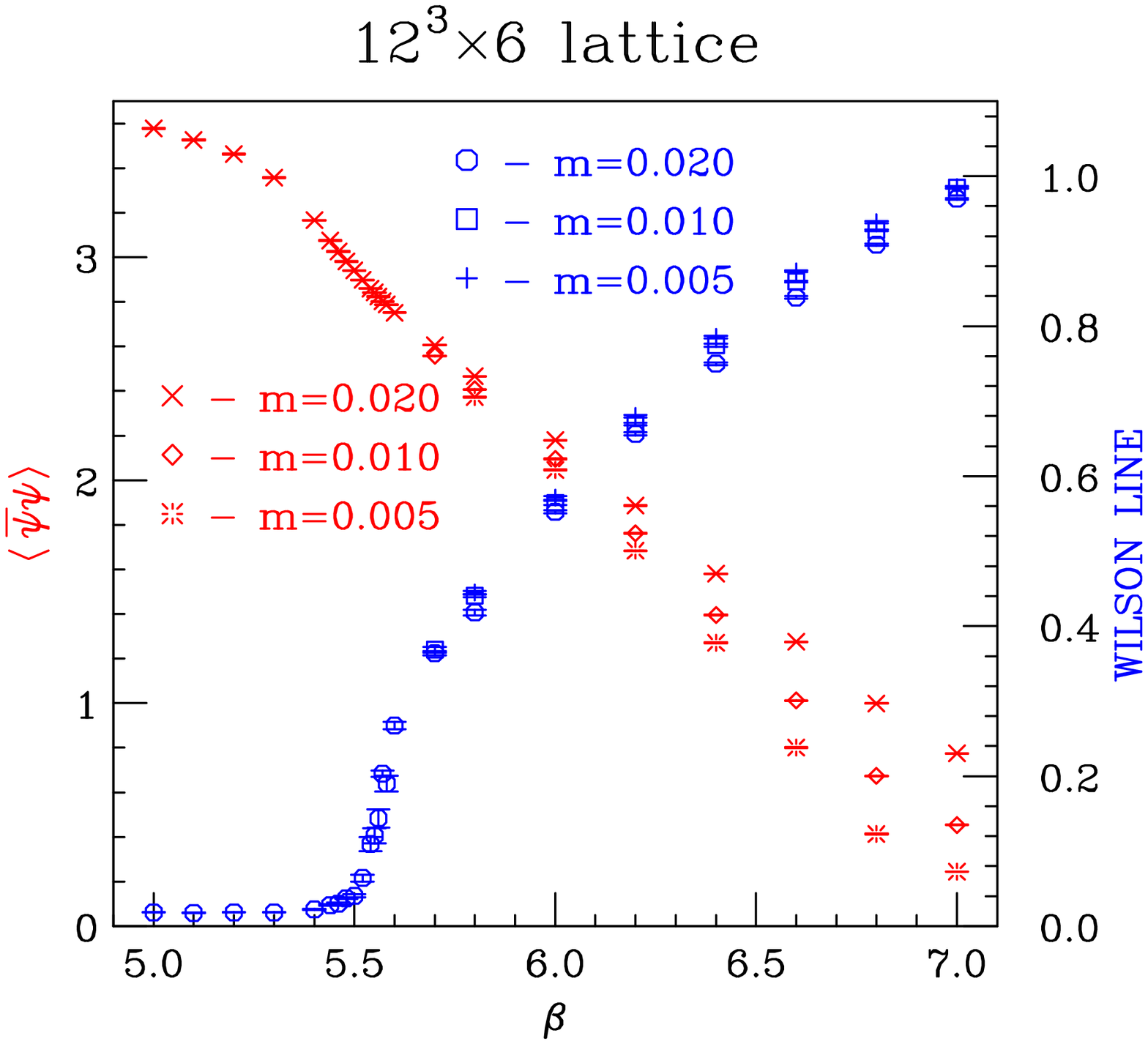}
\caption{Triplet Wilson Line and $\langle\bar{\psi}\psi\rangle$ as functions
of $\beta$ for the state with a real positive Wilson Line.}
\label{fig:wil-psi_12x6a}
}
\parbox{0.2in}{}
\parbox{2.9in}{
\epsfxsize=2.9in
\epsffile{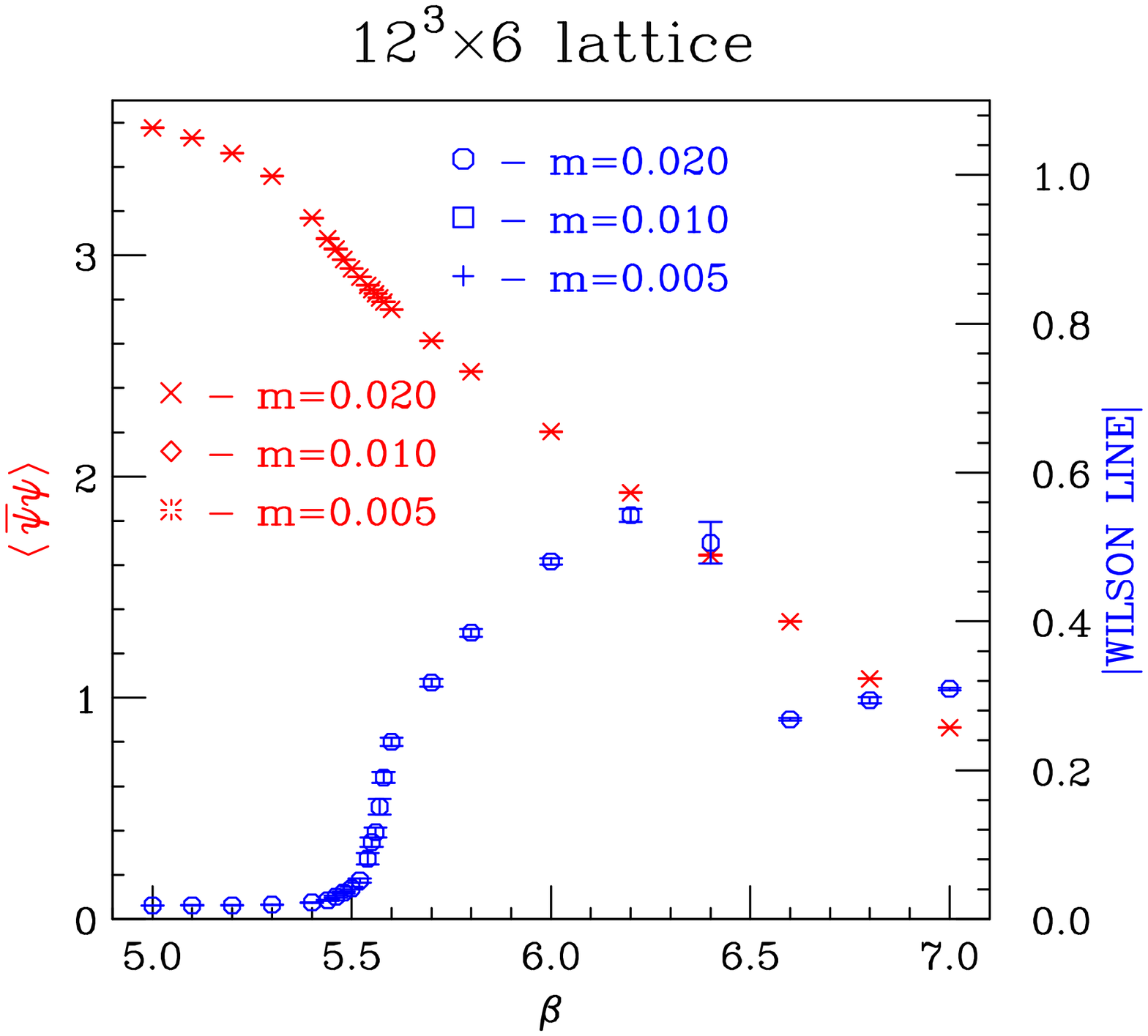}
\caption{Triplet Wilson Line and $\langle\bar{\psi}\psi\rangle$ as functions
of $\beta$ for states with complex or negative Wilson Line.}
\label{fig:wil-psi_12x6b}
}
\end{figure}

We bin our observables according to whether the arguments of the Wilson Lines
lie in the range $(-\pi,-\pi/3)$, $(-\pi/3,\pi/3)$, $(\pi/3,\pi)$. The `data'
in the first and last bins (corresponding to complex Wilson Lines) are combined,
complex conjugating where necessary. 
Figures~\ref{fig:wil-psi_12x6a},\ref{fig:wil-psi_12x6b} show the Wilson Lines
(Polyakov Loops) and the chiral condensates. The first graph is for the states
with real positive Wilson Lines. The second is for those with complex (or
negative) Wilson Lines.
From these graphs we conclude that the deconfinement transition for $m=0.02$
occurs at $\beta=5.54(4)$ and the chiral symmetry restoration transition is at
$\beta \approx 6.8$. The increase in the $\beta$s for each of these transitions
is what would be expected if they are finite temperature transitions for an
asymptotically free theory. We note that the fact that $Z_3$ symmetry is
broken by the quarks manifests itself in the difference in magnitudes of
the Wilson Line for the state with a positive and those with complex/negative 
Wilson Lines.

The states with complex Wilson lines show a transition to states with real
negative Wilson lines at $\beta \approx 6.5$. This increase over the value at
$N_t=4$ is much larger than for the deconfinement and chiral-symmetry
restoration transitions. This apparent inconsistency leads us to suspect
that this particular transition is a lattice artifact.

\section{Discussion and Conclusions}

In our studies of the thermodynamics of lattice QCD with 2-flavours of 
colour-sextet quarks we find well separated deconfinement and chiral-symmetry
restoration transitions. This contrasts with the situation with fundamental
quarks where these two transitions appear coincident, but is similar to the 
case of adjoint quarks where again these two transitions are separate.
\cite{Karsch:1998qj,Engels:2005te}

The increase in the $\beta$s for both transitions from $N_t=4$ to $N_t=6$ is
consistent with them being finite temperature transitions for an asymptotically
free theory rather than bulk transitions. If this theory has an infrared stable
fixed point, we have yet to observe it. Since we expect that $T_\chi \ge T_d$,
the fact that $\beta_\chi > \beta_d$ is also what would be expected for an
asymptotically free field theory. These very preliminary results would suggest
a walking rather than a conformal behaviour.

The phase structure we observe is very different from that observed with
Wilson quarks by DeGrand, Shamir and Svetitsky, who found coincident
deconfinement and chiral symmetry restoration transitions. It would be possible
for two different actions to give such different results if there is an infrared
fixed point (as these authors find) and we are on the strong-coupling side of
it. Alternatively, our quark masses might be too large to let us access the
chiral limit. Because of these differences it will be interesting when
simulations with sextet overlap quarks \cite{Fodor:2008hm} are extended to
larger lattices.

Let us summarise our preliminary results. At $N_t=4$, 
$\beta_d(m=0.02)=5.420(5)$, $\beta_d(m=0.01)=5.412(1)$; deconfinement appears
to be first order. $\beta_\chi \approx 6.5$. At $N_t=6$, 
$\beta_d(m=0.02)=5.54(4)$. $\beta_\chi \approx 6.8$.

For the deconfined phase there is a 3-state signal, the remnant of now-broken
$Z_3$ symmetry. For $N_t=4$ the states with complex Polyakov Loops appear
metastable. For $N_t=6$ all 3 states appear stable. Breaking of $Z_3$ symmetry
is seen in the magnitudes of the Wilson Lines for the states with positive real
versus those with complex/negative Wilson lines. The existence of the 3-state
signal is presumably because the formation of the chiral condensates at short
distances suppresses the contribution of the quarks at the confinement scale,
so the deconfined but chirally broken phase is more similar to the deconfined
phase of the quenched theory. (It would be interesting to compare this to
deconfined phase with heavy colour-triplet quarks).

Between the deconfinement and chiral transitions, we find a third transition
where the Wilson Lines in the directions of the 2 non-trivial roots of unity
change to real negative Wilson Lines. This transition occurs for 
$\beta \approx 5.9$ ($N_t=4$) and $\beta \approx 6.5$ ($N_t=6$). This rapid
increase suggests that the transition is a lattice artifact. If this third
transition is real, the fact that the magnitude of the negative Wilson line is
roughly one third of that for the positive Wilson line, suggests that this
transition might be associated with colour symmetry breaking 
$SU(3) \longrightarrow SU(2) \times U(1)$.

We need larger lattices -- $18^3 \times 6$ to study finite volume effects at
$N_t=6$ and $16^3 \times 8$ to see that the $N_t$ dependence of $\beta_d$ and
$\beta_\chi$ which we observe is not a lattice artifact of these coarse 
lattices. Smaller quark masses are needed to access the chiral limit 
unambiguously. To understand this theory more fully, we need to study its zero
temperature behaviour, measuring its spectrum, string tension, potential,
$f_\pi$... . Measurement of the running of the coupling constant for weak
coupling is needed.

It would be useful to repeat these simulations for 3 flavours of sextet quarks,
where it is believed that there should be an infrared stable fixed point, to
see if we find a qualitatively different phase diagram. Lattice simulations
should be applied to other candidate theories.

\section*{Acknowledgements}
These simulations were performed on the Cray XT4, Franklin at NERSC under an
ERCAP allocation and on the Linux cluster, Abe at NCSA under an LRAC grant.

\end{document}